\documentstyle[prl,aps,multicol,epsf]{revtex}

\begin{document}
\draft 
\title{Phase diagram of quantum capacitive junctions.}
\author{Thomas Strohm and Francisco Guinea}
\address{Instituto de Ciencia de Materiales. CSIC. \\
	 Cantoblanco. E-28049 Madrid. Spain.}
\date{\today}
\maketitle
\begin{abstract}
Mesoscopic capacitances, in the quantum regime, are described as a
quantum rotor coupled to a dissipative bath. We analyze the phase
diagram of this model, as function of the capacitance, the coupling to
the bath (that is, the conductance of the device), and the decay of
the correlation functions in the bath, i.\ e., the strength of the many
body effects within the leads.  By combining Monte-Carlo and analytical
techniques, we present the full phase diagram. The similarities and
differences with other models of quantum junctions are discussed.
\end{abstract}
\vskip0.2cm
\pacs{PACS numbers: 75.10.Jm, 75.10.Lp, 75.30.Ds.}

\begin{multicols}{2}

Quantum dissipative systems\cite{CL,Likharev,Weiss} are models for
mesoscopic electronic devices. They consist of one, or more, tunnel
junctions between normal or superconducting electrodes. Quantum
effects are associated to the dynamics of a collective variable which
describes the overall phase of the electronic wavefunction. For the
case of superconducting systems, this variable is the phase of the
order parameter\cite{AES}. A collective variable, conjugate to the
total charge, can be defined for normal junctions as
well\cite{GS,SZ}. The magnitude of the quantum fluctuations is
governed by the capacitances in the device.

Standard models describe these systems in terms of a few collective
variables. In the simplest case, these variables are the phase
mentioned previously, and its conjugate variable, the charge.  The
influence of the remaining, microscopic, degrees of freedom is taken
into account by means of effective transport coefficients (viscosity,
conductance) which couple the collective variable to a reservoir,
described in terms of harmonic oscillators. Two generic models have
been analyzed within this framework: i) The Caldeira-Leggett
model\cite{CL} which describes the dynamics of a junction where the
quantization of charge is unimportant, and ii) The quantum
rotor\cite{GS,Schoen}, in which only quantized charge transfer
processes take place. The first model has been extensively studied. A
generalization of it, the Schmid model\cite{Schmid}, also describes
junctions between Luttinger liquids\cite{KF}.

In the following, we will study in detail the second of these models.
We consider a system made of gapless components, in which quantum
(charging) effects are important. We do not include, in the
description of the environment, processes which destroy the charge
quantization in the device, such as an external electromagnetic field.
We consider the effective action:
\begin{eqnarray}
S_{\it eff} &= &\frac{\hbar}{2 E_C} \int  {\dot{\theta}}^2 d \tau \nonumber
\\ &+
&\alpha \tau_c^{2 g - 2} \int d \tau \int d \tau ' 
\frac{1 - \cos ( \theta_\tau - \theta_{\tau '} )}{( \tau - \tau ' 
)^{2 g}}
\label{action}
\end{eqnarray}

The model has four parameters: i) a short time cutoff, $\tau_c$,
ii) the charging energy,
$E_C = ( e^2 / C )^{-1}$, where 
$C$ is the capacitance of the device, iii) a
dimensionless coupling, $\alpha \sim \hbar / ( e^2 R )$, where $R$ is
the resistance, and iv) the exponent $g$.  The case of $g = 1$
corresponds to a quantum capacitance between two normal metals. $g \ne
1$ implies the existence of strong many body effects in the
electrodes. They may arise because of shake-up and exciton
effects\cite{UG} induced by electrons as they tunnel, closely related
to the Mahan-Nozi\`eres-de Dominicis\cite{MND} processes in X-ray
photoemission. An alternative way of inducing a value of $g \ne 1$ is
when the electrodes are one dimensional Luttinger liquids, a situation
which can be realized experimentally\cite{exp}.

The model given in (\ref{action}), has been studied extensively by a
variety of methods. A standard perturbative renormalization group
analysis can be performed when $\alpha \gg 1$ and $g \sim
1$\cite{Kosterlitz}.  The scaling shows the existence of a phase
transition along a critical line, $\alpha_c \propto ( 1 - g_c )^{-1}$.

The case $g = 1$ has also been analyzed
analytically\cite{BS1,Zwerger,PZ}, and
numerically\cite{BS2,Falci}. The system shows a crossover from a high
$\alpha$ regime, dominated by small amplitude oscillations (spin
waves), to a disordered phase for small values of $\alpha$. The
possibility of a sharp transition, instead of a crossover, has also
been postulated\cite{Zwerger,BS2,Falci}.  Because of the periodicity
in the variable $\theta$, different sectors, characterized by the
conjugate variable to $\theta$, can be defined. The edge of the
Brillouin Zone generated in this way can be mapped onto a dissipative
two level system\cite{GS,Falci2}, and solved.  When $E_C = 0$, the rotor
loses its dynamics. In terms of the environment, implicit in
(\ref{action}), we can define an effective Schmid model, with
a non linear potential acting on the position closest to the rotor
(see below).

The model is also related to the ``quantum rotor'' problem, introduced
in the study of monopole induced proton decay in grand unified field
theories\cite{Polchinski}. This model can be solved exactly,
in the limit $E_C \tau_c \rightarrow 0$,
(see\cite{Polchinski} and also\cite{conformal}).  As will be discussed
later, this model is equivalent to eq.~(\ref{action}) for $ \alpha = N$
and $g = 1$,
where $N$ is the number of fermion flavors coupled to the rotor.

We will study the action in eq.~(\ref{action}) by means of a numerical
Monte-Carlo procedure, supplemented by analytical methods, and by
comparing to the extensive literature mentioned above. The dynamics of
the collective variable $\theta$, arises from its interaction with the
microscopic degrees of freedom of the environment.  Instead of
studying numerically a 1D system with long range interactions, we
analyze a 2D model with short range interactions.  The environment is
modeled by a 2D X-Y system, with couplings $J_x = J_{\tau} = J_B$,
were the subscripts $x$ and $\tau$ stand for the space and the
imaginary time dimensions.  The planar spins are defined at the nodes
of a square lattice, with periodic boundary conditions. We assign to
each spin an angle, $\phi_{x,\tau}$.  In order for the environment to
show algebraic correlations, we have $J_B > J_c \approx 1.12$.  For
$J_B \gg J_c$, a standard spin wave analysis gives:
\begin{equation}
\langle \cos ( \phi_{x,\tau} - \phi_{x,\tau '} ) \rangle
\propto \frac{1}{( \tau - \tau ' )^{\frac{J_c}{ 4 J_B}}}
\label{correl}
\end{equation}

\begin{figure}
\narrowtext
\epsfxsize=3.1in
\hspace{.2em}
\epsffile{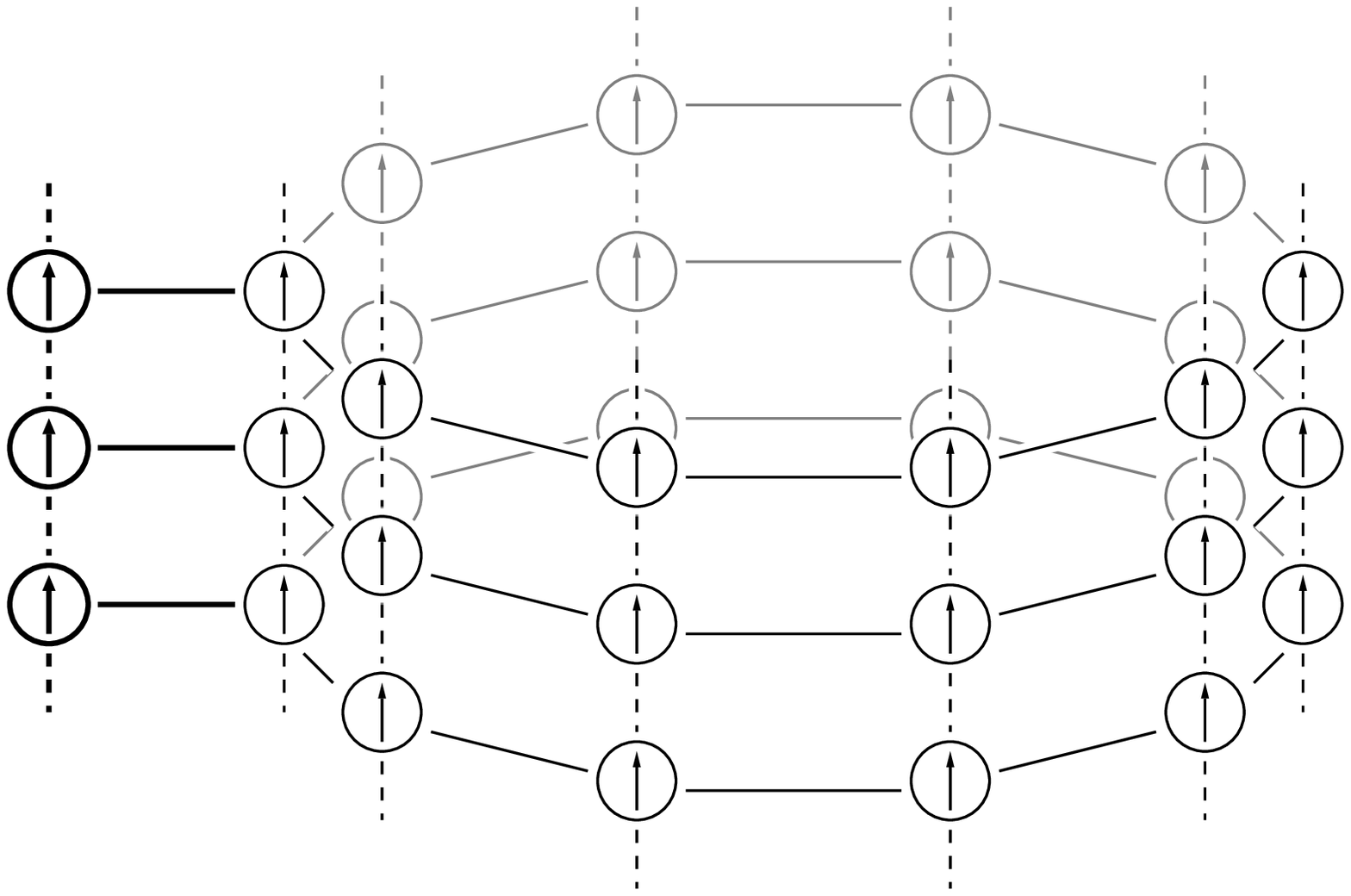}
\vspace{1.5\baselineskip}
\caption{Sketch of the model analyzed numerically in the text.}
\label{model}
\end{figure}
\vspace{0.5\baselineskip}

The rotor is defined as an additional 1D X-Y model which is coupled to
the environment in the way shown in fig.~(\ref{model}).  It introduces
two parameters, $J_R$ and $J_D$. The first one determines the
correlation length of the rotor when decoupled of the bath, and the
second gives the magnitude of the interaction with the environment. In
terms of the action (\ref{action}), $J_R \propto E_C^{-1}$, and,
approximately, $J_D^2 \propto \alpha$. The rotor is described in terms
of the angle $\theta$. We couple it to the nearest column of
environment spins by a term:
\begin{equation}
S_{\it coupling} = J_D \sum_{\tau} \cos ( \theta_{\tau} - n
\phi_{x_0 , \tau} )
\label{coupling}
\end{equation} 
where $n$ is an integer defined to facilitate the tuning of the
parameter $g$ in (\ref{action}).

By integrating out the environment degrees of freedom, we obtain an
effective action, which, at long imaginary times, behaves like
(\ref{action}).  The periodicity of the variables in our discrete model
constrains the effective interactions between the $\theta_{\tau}$'s to
be also periodic. The existence of algebraic decay in the environment
(\ref{correl}), determines a similar decay for the effective retarded
interactions of the rotor. An exact mapping of the parameters $J_B,
J_R, J_D$ into $\tau_c , \alpha , g$ can only be done numerically. A
reasonable approximation can be obtained by integrating out the
environment using second order perturbation theory, and
reexponentiating the result.  Then, as mentioned earlier, $\alpha \sim
J_D^2$. Finally, using the spin wave result (\ref{correl}), we find $2
g = ( n^2 J_c ) / ( 4 J_B )$. Note that, as $J_B > J_c \approx 1.12$, the only
way of achieving values of $2 g > 0.3$ is to make $n > 1$ (we will
mostly take $n = 4$).

\begin{figure}
\narrowtext
\epsfxsize=3.1in
\hspace{.2em}
\epsffile{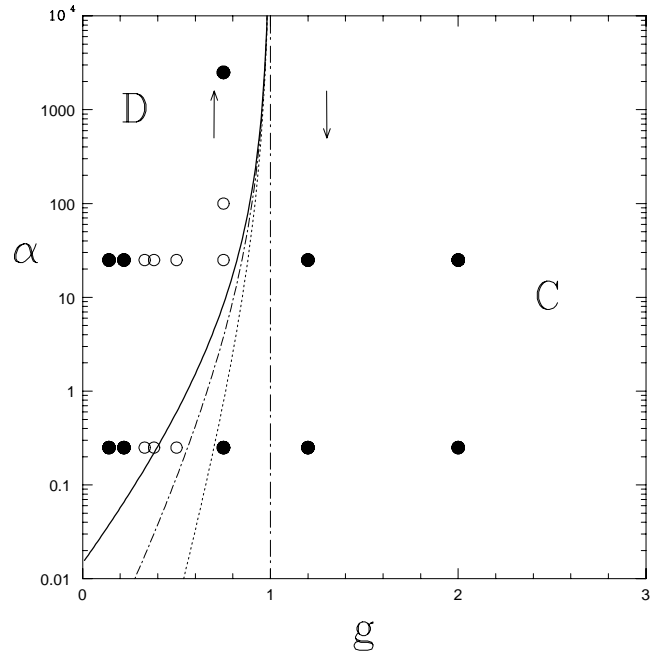}
\vspace{1.5\baselineskip}
\caption{Phase diagram of the model. Circles are
points analyzed by Monte-Carlo ($E_C = 0.4$, in umits of
the cutoff).  The full circles denote cases
when an
unambiguous characterization of the correlation functions can be
made. The lines denote the prediction of the critical line from
the variational spin wave analysis discussed in the text.
Full line, $E_C = 0.4$. Broken line, $E_C = 0.2$. Dashed line,
$E_C = 0.05$.
Arrows in the top part give the flow of
the RG scaling equations for $\alpha \gg 1$ and $g \sim 1$. 
The two phases that we
find are denoted C (capacitive) and D (dissipative).} 
\label{diagram}
\end{figure}
\vspace{0.5\baselineskip}

We have studied the model by Monte-Carlo\cite{Montecarlo}. Typical
system sizes are $l_x = 20 - 40$ and $l_{\tau} = 100 - 200$. The
number of configurations used is $\sim 20000$.  The samples, 
produced by the Metropolis algorithm, were not
correlated (autocorrelation $< 0.1$). The cases analyzed are shown in
fig.~(\ref{diagram}). The two phases depicted in this figure are
characterized by the rotor correlation functions:
\begin{equation}
G ( \tau - \tau ' ) = \langle \cos ( \theta_{\tau} - \theta_{\tau '} )
\rangle
\label{green}
\end{equation}

\begin{figure}
\narrowtext
\epsfxsize=3.1in
\hspace{.2em}
\epsffile{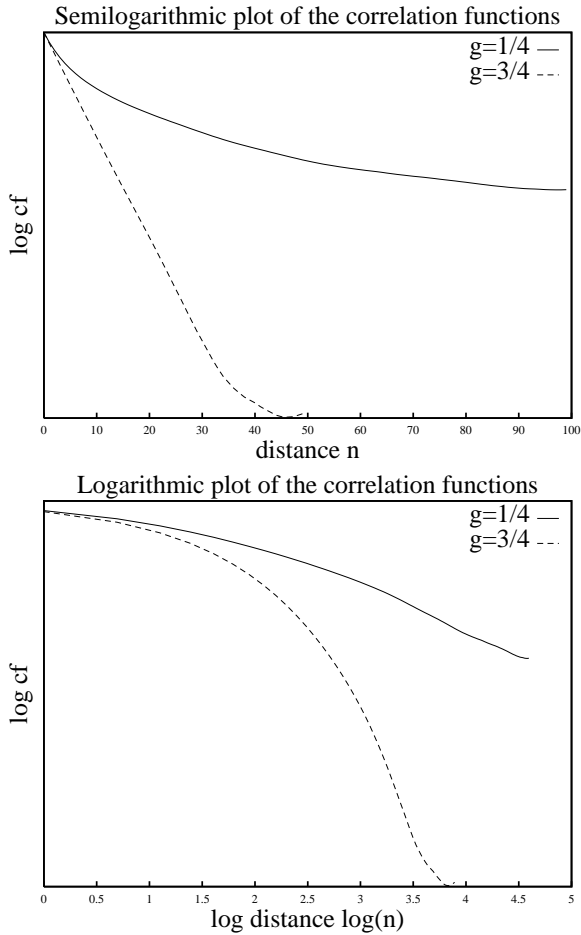}
\vspace{1.5\baselineskip}
\caption{Values of $\langle \cos ( \tau - \tau ' ) \rangle$ as
function of $\tau - \tau '$. The two cases shown correspond to $g = 1
/ 4$ and $\alpha \sim 0.27$ ($J_D / J_c = 0.5$), full line, and $g = 3
/ 4$ and $\alpha \sim 0.27$, broken line.}
\label{greenf}
\end{figure}

This function shows two distinct regimes, as shown in
fig.~(\ref{greenf}). For sufficiently large values of $g$, or low
values of $\alpha$, the correlation (\ref{green}) decays
exponentially, while, if $g$ is small, it shows an algebraic
dependence on $\tau - \tau '$.

The thick line shown in fig.~(\ref{diagram}) separates the two
regimes.  In terms of the original model, an algebraic decay in
(\ref{green}) implies dissipative behavior, although it needs not be
ohmic.  When (\ref{green}) decays exponentially, the system behaves as
a capacitor, with an effective capacitance proportional to the inverse
correlation length\cite{gap}.

\vspace{0.5\baselineskip}

We now study the phase diagram of the model by variational methods.
The standard way is to employ the self consistent harmonic
approximation (SCHA), which amounts to a spin wave approximation. The
couplings of the model are replaced by quadratic interactions, whose
strength is then optimized. This scheme has already been applied to
the action (\ref{action}), for $g = 1$\cite{BS1}, and at finite
temperatures. For the related Schmid model, it reproduces correctly
the phase diagram\cite{SCHA}.

For that purpose,  we use the hamiltonian formalism:
\begin{eqnarray}
{\cal H} &= &E_C \frac{P^2}{2} + V \left[ 1 - \cos  
\left( \theta - \phi_0 \right)
\right] + \nonumber \\
& &{} + \omega_c \sum_{n = 0}^{\infty} \frac{p_n^2}{2} + 
\frac{v_s^2 ( \phi_n - \phi_{n+1} )^2}{2}
\label{hamil}
\end{eqnarray}
where $[ \theta , P ] = [ \phi_n , p_n ] = i$ and $\alpha \propto ( V
/ \omega_c )^2$.  For a decoupled chain ($V = 0$), we have $ \lim_{t
\rightarrow \infty} \langle [ \phi_0 ( t ) - \phi_0 ( 0 ) ]^2 \rangle
= 1 / ( 4 \pi v_s ) \log ( \omega_c t )$, so that $2 g = 1 / ( 4 \pi
v_s )$. Note that, in this representation, the connexion of this model
and the Schmid model in the limit $E_C \rightarrow 0$ is transparent.

The SCHA hamiltonian is:
\begin{equation}
{\cal H}_{\it SCHA} = E_C \frac{P^2}{2} + E_K \frac{( \theta - \phi_0
)^2}{2} + {\cal H}_{\it chain}
\label{SCHA}
\end{equation}
where ${\cal H}_{\it chain}$ is the last part in eq.~(\ref{hamil}) and
$E_K$ is the variational parameter. We need to minimize $\langle {\it SCHA}
| {\cal H} | {\it SCHA} \rangle$, where $| {\it SCHA} \rangle$ is the ground 
state of ${\cal H}_{\it SCHA}$. In the limit $E_K , E_C \ll \omega_c$, we have:
\begin{equation}
\langle {\it SCHA} | {\cal H} | {\it SCHA} \rangle = c \sqrt{ E_C E_K}
- V \left( \frac{\sqrt{E_C E_K}}{\omega_c} \right)^{g}
e^{-\sqrt{\frac{E_C}{E_K}}}
\label{minimum}
\end{equation}
where $c$ is a numerical constant of order unity.

Eq. (\ref{minimum}) admits solutions for $g < 1$ and, roughly,
$\alpha > c / ( 1 - g ) ( E_C / \omega_c )^{(2 - 2 g )}$, where
$c$ is a numerical constant of order unity. The factor $( 1 - g )^{-1}$
has been inserted in order to interpolate correctly to the 
$\alpha \rightarrow \infty$, $g \rightarrow 1$ limit. This line
is plotted in fig.(\ref{diagram}).

The possibility of a description of the physics of the model in terms
of only spin waves does not necessarily imply that other effects are
irrelevant. Experience with related models, however, like the Schmid
model\cite{SCHA}, or the 1D quantum sine-Gordon gives some confidence
on the reliability of the approach.

We now turn to the relation of our work to the problem of
monopole induced baryon decay. The main issue addrssed in this context
has been the dynamics of the baryons which build up the environment.
It is, however, possible, to derive an effective action for
the rotor coordinate. The leading terms should be those given
in(\ref{action}). As the rotor is coupled to free fermions, 
$g = 1$. The analysis in the literature\cite{Polchinski,conformal},
initially neglects the periodicity in the rotor coordinate. 
It implies, in expression (\ref{action}), the substitution of
$1 - \cos ( \theta_{\tau} - \theta_{\tau '} )$ by
$ ( \theta_{\tau} -\theta_{\tau '} )^2 / 2$. In\cite{Polchinski}, 
the coupling (that is, the fermion charge)
 is included in the definition of $E_C$, and the value
of $\alpha$ is simply $N$, the number of fermion channels coupled
to the rotor.
This model reduces to
the Caldeira-Leggett model of a quantum particle in a dissipative
environment\cite{CL}, which can be solved exactly.  The rotor coordinate
experiences a frictional force, in agreement with the results in
\cite{Polchinski}.

The solutions obtained within this scheme describe the model in
terms of low energy electron hole pairs (spin waves, in our language).
Such a description exhaust the physics of the problem in the right
side of the phase diagram in fig.(\ref{diagram}). Thus, the validity
of the scheme depends on the value of $E_C / \omega_c$. It
is assumed in\cite{Polchinski,conformal} that $E_C / \omega_c 
\rightarrow 0$, so that a spin wave description is marginally
correct for $g = 1$.

Note that a wavefunction which restores the symmetry of the model
can be written by superposing  $| SCHA \rangle$ and the solutions of
${\cal H}_n = {\cal H}_{SCHA} + 2 \pi n E_K ( \theta - \phi_0 )$.
Such state, $\sum_n e^{i k n} | {SCHA}_n \rangle$, has the
same physical properties as $| SHA \rangle$, because
$\langle {SCHA}_n | {SCHA}_{n '} \rangle = \delta_{n , n'}$.

It is interesting to speculate what would happen if $E_c / \omega_c$
is finite. Then, quantum fluctuations in the rotor coordinate
would decouple it, at low energies, from the fermions.
The rate of monopole induced baryon decay would not be of order unity,
but would tend to zero as $\omega^2 / E_C^2$\cite{UG}.

Summarizing, we have studied the phase diagram of a model used to
describe a variety of physical systems, such as Coulomb blockade in a
normal metal junction, the single electron transistor or the quantum
box.  The generalization that we make in letting the parameter $g$ in
eq.~(\ref{action}) be different from 1 allows us to study other
situations, like charging effects in Luttinger liquid junctions, or
shake-up and excitonic processes in normal junctions. 

We find a phase diagram with two regions: one in which charging
effects dominate and the system behaves as a capacitor at low
energies, and another in which the quantization of the charge is
irrelevant, and the device shows a dissipative response.  This phase
diagram has been obtained by Monte-Carlo methods, and its consistency
checked against alternative analytical approaches, such as the RG for
$\alpha \gg 1$ and $g \sim 1$ and variational ans\"atze for $\alpha \ll
1$.
The phase diagram shows interesting deviations from the well
understood phase diagram of the Schmid model: i) there is no duality,
and the separatrix is not vertical, ii) there is a dynamically
generated scale below which gaplike features appear (in the insulating
phase) and iii) the charge quantization explicitly imposed in
eq.~(\ref{action}) reduces the the range of parameters for which the
system shows dissipative behavior.

We have studied the ground state only, and we have not considered the
influence of finite voltages, which are important in some situations,
like the single electron transistor. We can anticipate, however, that
gate or bias voltages will only be relevant for the phase shown to the
right in fig.~(\ref{diagram}).

One of us (T.~S.) thanks the Deutscher Akademischer Austauschdienst
for financial support. This work has also been supported by the CICyT
(Spain) through grant MAT94-0982.

\end{multicols}

\end{document}